\pdfoutput=1

\documentclass[11pt]{article}
\usepackage[final]{acl}
\usepackage{times}
\usepackage{latexsym}
\usepackage[T1]{fontenc}
\usepackage[utf8]{inputenc}
\usepackage{microtype}
\usepackage{inconsolata}

\usepackage{graphicx}
\usepackage{booktabs}
\usepackage{multirow}
\usepackage{xcolor}
\usepackage{colortbl}
\usepackage{amsmath}
\usepackage{amssymb}
\usepackage{pifont}
\usepackage{subcaption}
\usepackage{enumitem}
\usepackage{hyperref}
\usepackage{cuted}
\usepackage{caption}

\newcommand{\oeg}{Observable Evidence Grounding}
\newcommand{\icr}{Interaction Contingency Reasoning}
\newcommand{\srm}{Support-Response Modeling}
\newcommand{\OEG}{OEG}
\newcommand{\ICR}{ICR}
\newcommand{\SRM}{SRM}
\newcommand{\oegtab}{\shortstack{Observable Evidence\\Grounding}}
\newcommand{\icrtab}{\shortstack{Interaction Contingency\\Reasoning}}
\newcommand{\srmtab}{\shortstack{Support-Response\\Modeling}}
\newcommand{\full}{Video{+}ASR}

\newcommand{\cm}{\checkmark}
\newcommand{\xm}{\textcolor{black!55}{\ding{55}}}

\title{VIRA-Bench: Diagnosing Pedagogical Contingency Understanding\\
in Online Tutoring Videos}

\author{Weihang Wang\textsuperscript{1} \quad Hansu Gu\textsuperscript{2} \quad
Guanming Liu\textsuperscript{1} \quad
Peng Zhang\textsuperscript{1*} \quad
Ning Gu\textsuperscript{1} \quad Tun Lu\textsuperscript{1}\thanks{Corresponding authors.} \\
\textsuperscript{1}Fudan University, Shanghai, China \quad
\textsuperscript{2}Independent Researcher, Seattle, United States \\
\texttt{whwang24@m.fudan.edu.cn \quad hansug@acm.org \quad gmliu24@m.fudan.edu.cn} \\
\texttt{zhangpeng\_@fudan.edu.cn \quad ninggu@fudan.edu.cn \quad lutun@fudan.edu.cn}}

\begin{document}
  \maketitle

  \setlength{\stripsep}{0pt}
  \begin{strip}
    \centering
    \includegraphics[width=\textwidth]{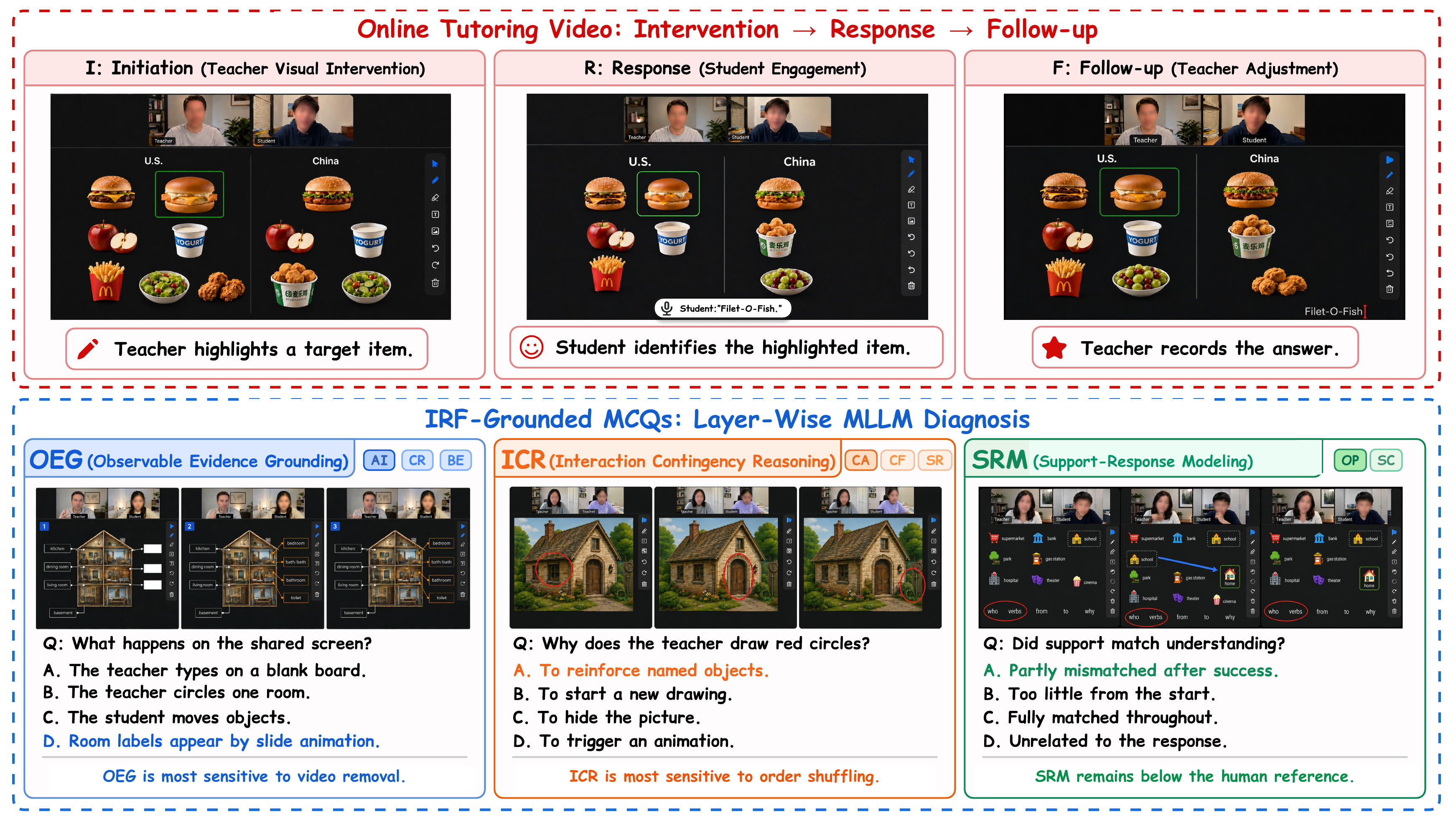}
    \captionof{figure}{VIRA-Bench overview. An online tutoring clip is organized as an IRF exchange and instantiated as IRF-grounded MCQs across OEG, ICR, and SRM. The controls reveal layer-specific evidence dependence: video removal affects OEG most, order shuffling affects ICR most, while SRM remains below the human reference subset.}
    \label{fig:teaser}
  \end{strip}

  \begin{abstract}
  \vspace{-8pt}
  Understanding a one-to-one tutoring video requires more than recognizing its
  events. A model must connect a visible teacher intervention to the learner
  response it elicits and to the follow-up adjusted to that response. Existing
  video and educational-dialogue benchmarks typically emphasize event
  recognition, temporal reasoning, social-state inference, or transcript-based
  interaction, but rarely diagnose this action--response--adjustment relation.
  We introduce \textbf{VIRA-Bench}, a focused diagnostic benchmark for
  short-horizon pedagogical contingency in one-to-one online tutoring.
  VIRA-Bench contains 1{,}005 tutoring clips and 1{,}322 multiple-choice
  questions with researcher evidence audit and expert verification, organized
  into \oeg{} (OEG), \icr{} (ICR), and \srm{} (SRM). Across 11 models, the best
  reaches 61.3\%, well below the 91.0\% human reference on a stratified subset,
  and ICR generally has the lowest layer accuracy among leading models. Under
  controlled input manipulations, removing video most strongly degrades OEG,
  whereas disrupting temporal order most strongly degrades ICR. Free-response,
  \textsc{Stem+Options}, and synchronized-shuffle controls show that response
  format, static question cues, and video--ASR mismatch do not explain these
  layer-specific patterns. The results identify distinct behavioral evidence
  dependencies without claiming an internal architectural mechanism.
  VIRA-Bench provides a focused evaluation resource for multimodal dialogue
  grounding, pedagogical discourse, and temporal reasoning in online tutoring.
\end{abstract}

  \section{Introduction}
\label{sec:intro}

Understanding a tutoring video requires more than recognizing its events. When a
teacher highlights a word, asks a question, and reacts to the learner's reply, a
model must align language with visible actions, recover event order, identify
learner uptake, and determine whether the follow-up is conditioned on that
response. This multimodal NLP problem spans dialogue grounding, discourse
structure, and interactional pragmatics. We refer to this focused capability as
understanding \textbf{visually grounded pedagogical contingencies}.

MLLM evaluation has expanded from visual perception and question answering to
long-form video comprehension, fine-grained temporal reasoning, and social
interaction reasoning
\citep{liu2023llava,Chen2023InternVS,Zhang2024LLaVAVideoVI,Yin2023ASO,Li2023MVBenchAC,Fu2024VideoMMETF,Mangalam2023EgoSchemaAD,Liu2024TempCompassDV,Cai2024TemporalBenchBF,Kong2025SIVBenchAV}.
These benchmarks primarily examine event recognition, temporal relation recovery,
or holistic social-state inference. Educational-dialogue resources model teacher
and learner turns but typically rely on transcripts
\citep{Demszky2022TheNT,Macina2023MathDialAD,Macina2025MathTutorBenchAB}.
Neither line jointly tests whether a model can connect a visible intervention to
the response it elicits and then connect that response to a subsequent
adjustment. A model may therefore recognize each action and its order without
recovering the interactional contingency that links them.

Online tutoring provides a natural setting for evaluating this capability.
Grounded in the zone of proximal development \citep{Vygotski1978MindIS},
formative assessment \citep{Black1998AssessmentAC,RuizPrimo2007ExploringTI}, and
contingent scaffolding \citep{Hattie2007ThePO,Pol2015TheEO}, effective
instruction requires detecting learner understanding and adjusting support in
real time. In tutoring interactions, teachers perform visible operations on a
shared screen (circling, highlighting, dragging); learners respond, and teachers
confirm or correct. A short clip therefore exposes an action, response, and adjustment
chain with explicit visual and temporal evidence.

Table~\ref{tab:gap} uses five binary criteria. Temporal video requires ordered
evidence; pedagogical interaction links teaching, learner response, and follow-up;
theory anchor denotes an explicit theoretical basis; diagnostic layers separate
cognitive operations; and temporal-order perturbation tests event-order dependence.
Existing resources do not combine all five.

\begin{table*}[t]
  \centering
  \footnotesize
  \begin{tabular*}{\textwidth}{@{\extracolsep{\fill}}l ccccc@{}}
    \toprule                                                 & \textbf{Temporal} & \textbf{Pedagogical} & \textbf{Theory} & \textbf{Diagnostic} & \textbf{Temporal-order} \\
                                                             & \textbf{video}    & \textbf{interaction} & \textbf{anchor} & \textbf{layers}     & \textbf{perturbation}   \\
    \midrule \multicolumn{6}{l}{\textit{Video Understanding}} \\[2pt]
    MVBench~\citep{Li2023MVBenchAC}                           & \cm               & \xm                  & \xm             & \cm                 & \xm                     \\
    Video-MME~\citep{Fu2024VideoMMETF}                       & \cm               & \xm                  & \xm             & \xm                 & \xm                     \\
    EgoSchema~\citep{Mangalam2023EgoSchemaAD}                & \cm               & \xm                  & \xm             & \xm                 & \xm                     \\
    SIV-Bench~\citep{Kong2025SIVBenchAV}                     & \cm               & \cm                  & \cm             & \cm                 & \xm                     \\
    Social IQa~\citep{Zadeh2019SocialIQAQ}                   & \xm               & \cm                  & \xm             & \xm                 & \xm                     \\
    \midrule \multicolumn{6}{l}{\textit{Educational NLP}}     \\[2pt]
    NCTE Transcripts~\citep{Demszky2022TheNT}                & \xm               & \cm                  & \cm             & \cm                 & \xm                     \\
    MathDial~\citep{Macina2023MathDialAD}                    & \xm               & \cm                  & \xm             & \xm                 & \xm                     \\
    MathTutorBench~\citep{Macina2025MathTutorBenchAB}        & \xm               & \cm                  & \cm             & \cm                 & \xm                     \\
    \midrule \textbf{VIRA-Bench (Ours)}                      & \cm               & \cm                  & \cm             & \cm                 & \cm                     \\
    \bottomrule
  \end{tabular*}
  \caption{Comparison with video understanding and educational interaction
  benchmarks.}
  \label{tab:gap}
\end{table*}

We present \textbf{VIRA-Bench}, a diagnostic benchmark for evaluating whether
video-language models can understand short-horizon pedagogical
contingencies, instantiated through one-to-one online tutoring interactions. The design
draws on two theoretical foundations. The Initiation, Response, and Follow-up (IRF)
structure \citep{Sinclair1975TowardsAA} characterizes local classroom exchanges
as three-part sequences. Contingent teaching theory \citep{Pol2010UvADAREA,Wood1976TheRO}
describes instructional support as adapting to evidence of learner understanding.
Based on these foundations, VIRA-Bench decomposes visually grounded
interactional contingency understanding into three diagnostic layers
distinguished by the cognitive operation required. \textbf{\oeg{}} tests whether
models can ground answers in observable actions and screen content. \textbf{\icr{}}
tests whether models can recover directed relations between observations, such as
linking an intervention to the learner response it elicits. \textbf{\srm{}} tests
whether models can predict learner responses and judge whether the support level matches
the learner state. The benchmark contains 1{,}005 tutoring clips (15--60\,s each)
and 1{,}322 multiple-choice questions across eight diagnostic subtasks,
constructed through a multistage pipeline where LLMs generate candidate
descriptions and questions, researchers audit evidence against source videos,
and domain experts verify a stratified item sample.

We evaluate 11 commercial and open-source MLLMs under the full Video+ASR setting
and analyze five representative models with ASR-only and temporally shuffled
inputs. Overall accuracy ranges from 20.3\% to 61.3\%, leaving a 29.7-point gap
between the best model and the 91.0\% human reference on a stratified subset.
Removing video reduces \oeg{} accuracy by 12.7--22.0 points, whereas disrupting
visual event order reduces \icr{} accuracy by 11.9--19.1 points. Free-response,
\textsc{Stem+Options}, and synchronized-shuffle controls further probe dependencies
on response format, static question cues, and video--ASR alignment.
Together, these results show that the three layers expose distinct behavioral
dependencies in visual grounding and temporal-order reasoning rather than
collapsing into a single video-QA score.

Our contributions are:
\begin{enumerate}[nosep, leftmargin=*]
  \item We formulate visually grounded pedagogical contingency understanding as
    a diagnostic evaluation problem for video-language models and instantiate it through
    theory-grounded one-to-one online tutoring interactions.

  \item We construct \textbf{VIRA-Bench}, which comprises 1{,}005 tutoring clips and
    1{,}322 MCQs across three diagnostic layers and eight subtasks, with LLM-based
    generation, researcher evidence audit, and expert item verification.

  \item We evaluate 11 MLLMs and show, through paired evidence interventions and
    response-format controls, that the three layers exhibit distinct dependencies
    on visual content and event order.
\end{enumerate}

  \section{Related Work}
\label{sec:related}

\paragraph{Video benchmarks.}
Video benchmarks now cover short-form perception, long-form understanding, and domain-specific challenges \citep{Li2023MVBenchAC,Fu2024VideoMMETF,Mangalam2023EgoSchemaAD,Ataallah2024InfiniBenchAC,Ray2025EduVidQAGA,Shafique2025ACM}.
Diagnostic suites further test reliance on temporal evidence or static shortcuts \citep{Liu2024TempCompassDV,Cores2024LostIT,Shangguan2024TOMATOAV,Cai2024TemporalBenchBF,Yang2025AudiocentricVU}, motivating controlled input conditions.
Adjacent work studies LLM reasoning and agents \citep{wu-etal-2026-beyond-examples,wu2026seed,wu-etal-2026-spark,yang2026opid,liu2026streammembench,liu2026intpro}, and calibration, hallucination, context optimization, and culture-dependent multimodal interpretation \citep{chen2024dynamic,wang2025diving,yang2026mememind,wang2026memebench}.
Causal video QA benchmarks, however, still emphasize physical or event causality rather than response-dependent participant contingency \citep{Xiao2021NExTQANP,Wu2024STARAB,Yi2019CLEVRERCE}; Appendix~\ref{app:causal_positioning} provides a detailed comparison.

\paragraph{Educational dialogue and scaffolding theory.}
Recent work has begun studying social and educational interaction in video \citep{Kong2025SIVBenchAV,Li2025MimeQATS,Selvakumar2025MULTIVOXAB}.
In educational NLP, \citet{Demszky2021MeasuringCU} operationalize conversational uptake, the NCTE Transcripts \citep{Demszky2022TheNT} provide large-scale classroom dialogue data, \citet{Demszky2023MPoweringTN} show NLP-powered feedback improves instruction, and MathDial \citep{Macina2023MathDialAD} and MathTutorBench \citep{Macina2025MathTutorBenchAB} evaluate LLM tutoring with pedagogically grounded criteria.
\citet{Demszky2026MappingTM} identify the gap between transcript-based NLP and video-based observation.
Automated classroom observation tools such as CLASS \citep{Pianta2008ClassroomAS} have been adapted with LLMs \citep{whitehill2024automatedevaluationclassroominstructional,Wang2023IsCA}, and ClassMind \citep{Qu2025ClassMindSC} extends this to video-based feedback, though on pre-segmented lessons.
\citet{Dou2025SimulatorArenaAU} show that user simulators can proxy multi-turn evaluation, highlighting the need for complementary benchmarks that directly test interaction understanding from recorded evidence.
VIRA-Bench addresses this gap by translating the IRF exchange structure \citep{Sinclair1975TowardsAA} and contingent scaffolding theory \citep{Wood1976TheRO,Pol2010UvADAREA} into a computational benchmark with paired evidence conditions.

  \begin{figure*}[t!]
\centering
\includegraphics[width=\textwidth]{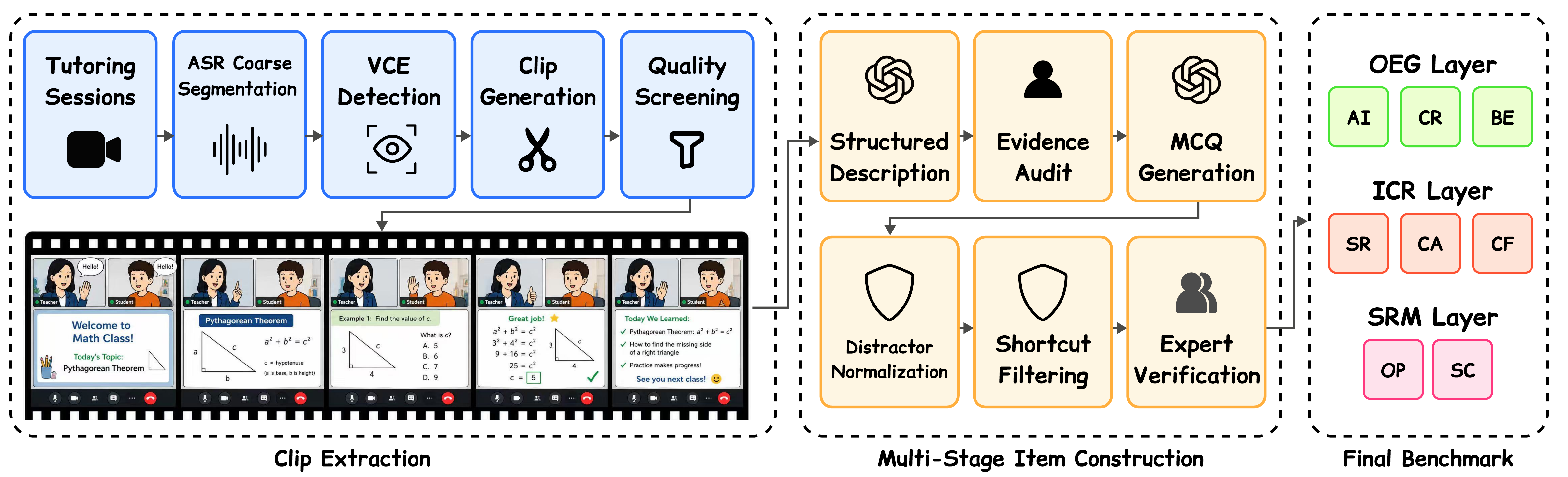}
\caption{VIRA-Bench construction pipeline. Clip extraction combines ASR-guided segmentation, Visual Content Event detection, and quality screening. Question construction uses separate LLMs for evidence description, MCQ generation, shortcut filtering, and difficulty calibration.}
\label{fig:pipeline}
\end{figure*}

\section{The VIRA-Bench Benchmark}
\label{sec:benchmark}

\subsection{Task Formulation}
\label{sec:task_def}

VIRA-Bench evaluates whether video MLLMs can understand short-horizon pedagogical contingencies in one-to-one online tutoring.
Given a 15--60 second tutoring clip containing a shared screen and a student webcam feed, a model answers multiple-choice questions about the interaction.
The central construct is \textbf{visually grounded interactional contingency}: a short-horizon exchange in which a teacher intervention elicits a learner response and the teacher then follows up based on that response, all inspectable from the clip.
Routine responses emerge within seconds, while cognitively demanding responses may take longer \citep{Rowe1986WaitTS}; VIRA-Bench focuses on local contingency within a single exchange rather than longitudinal scaffolding fading across sessions.

\subsection{Diagnostic Framework}
\label{sec:framework}

We design a three-layer diagnostic framework grounded in the IRF exchange structure and the contingent teaching cycle.
The layers are separated by the \emph{type of cognitive operation} required \citep{Chi2014TheIF}, rather than by a simple difficulty ordering.
In \oeg{}, models must ground the interaction in observable evidence from the screen, teacher action, and student behavior.
In \icr{}, they must connect these observations through temporal and contingent relations, asking whether the teacher's intervention plausibly triggered the student's response and whether the teacher's follow-up targeted that response.
In \srm{}, they must model the support--response relation that the exchange implies: predicting how a learner is likely to respond to a teacher's support move, or judging whether that support level matches the learner state revealed by the response.

The eight subtasks operationalize these three cognitive operations while keeping each question targeted at one primary capability.
For example, a causal attribution question may show a teacher circling a word on the shared screen, followed by the student typing a response in the chat box.
The model must distinguish whether the student's response follows from the visual intervention, precedes it, merely coincides with it, or reflects an unsupported alternative explanation.
Such questions test whether the model can use temporal evidence to recover a plausible I$\to$R contingency, rather than only recognizing isolated events.

\subsection{Clip Extraction from Tutoring Sessions}
\label{sec:data}

Source videos are drawn from a corpus of over 1{,}000 professionally recorded online English tutoring sessions, each lasting 1--2 hours, collected through video-conferencing platforms between 2018 and 2021.
The released benchmark spans 72 learners and 22 teachers at primary and secondary levels.
Most learners speak Mandarin as their first language, while a small number are native English speakers.
Teacher speech therefore naturally includes code-switching between English instruction and Mandarin clarification.
All faces are blurred to protect participant privacy, with empirically verified impact on pedagogical content visibility below 3\%.

Clips are extracted through a coarse-to-fine process.
First, ASR-based coarse segmentation identifies candidate instructional intervals, reducing each 1--2 hour session to approximately 15--30 minutes of likely interaction.
Second, Visual Content Event (VCE) detection compares consecutive shared-screen frames using structural similarity \citep{Wang2004SSIM} to locate moments where the teacher produces a visual intervention, such as writing, circling, highlighting, or dragging an object.
Third, each VCE is used as an anchor point for clip generation, yielding a 15--60 second window centered on a single teacher intervention and its surrounding interactional context.

The clip-level pipeline starts from 2{,}760 VCE anchors.
After merging nearby anchors and removing duplicate windows, 1{,}840 candidate clips remain, filtering 33.3\% of the anchors.
Clip quality screening further removes 19.6\% of candidates whose shared screen is unclear, whose student response window is missing, or whose interaction contains multiple overlapping exchanges.
This yields 1{,}480 clips.
After structured evidence description and researcher audit, 1{,}370 clips remain, with 7.4\% removed due to unrecoverable missing or inconsistent evidence.
Finally, question-level filtering keeps 1{,}005 clips with at least one valid MCQ, filtering 26.6\% of clips at this stage.
Extraction parameters are reported in Appendix~\ref{app:extraction_params}.

\subsection{Multi-Stage Question Construction}
\label{sec:pipeline}

VIRA-Bench follows a construction--evaluation separation protocol with two main stages. First, each clip is converted into an audited structured description recording the observable and temporal evidence needed for pedagogical interpretation. Second, candidate MCQs are generated from these descriptions and filtered through model-based checks and expert audit. To reduce direct construction--evaluation overlap, the two stages use disjoint model sets: GPT-4o \citep{openai2024gpt4o} produces structured descriptions and GPT-5 \citep{openai2025gpt5} generates candidate MCQs, while held-out validators GLM-5V-Turbo \citep{zhipu2026glm5v} and Claude-Haiku-4.5 \citep{anthropic2025claudehaiku} handle shortcut filtering and diagnostic calibration. All construction models are excluded from the evaluation leaderboard.

\subsubsection{Structured Evidence Description}
\label{sec:evidence_description}

GPT-4o receives each clip as video input together with its ASR transcript and produces a structured JSON description covering five evidence dimensions: (1)~\emph{screen events}, including timestamped teacher actions on the shared screen such as drawing, highlighting, dragging, and slide transitions; (2)~\emph{student behaviors}, including observable verbal responses, gestures, gaze shifts, and screen interactions; (3)~\emph{teacher speech acts}, including ASR-derived utterances categorized by pragmatic function such as instruction, question, feedback, and elicitation; (4)~\emph{temporal dynamics}, including the intervention--response pattern and estimated response latency; and (5)~\emph{cognitive demand level}, describing the complexity of the task posed by the teacher.

These descriptions are treated as evidence drafts rather than final annotations. Before question generation, researchers audit each description against the source video, correct factual errors, and remove clips whose key events cannot be reliably recovered. Only audited descriptions are used for MCQ generation. This step provides an explicit evidence interface between raw videos and question construction, reducing the risk that generated items depend on unsupported interpretations.

\subsubsection{Question Construction and Filtering}
\label{sec:mcq_generation}

GPT-5 generates candidate questions from the audited descriptions, with the target subtask specified in the prompt. Distractors are produced in a separate pass under three normalization constraints that reduce superficial answer cues: (1)~\emph{length matching}, where all four options must fall within 20\% of the mean option length; (2)~\emph{specificity matching}, where distractors must refer to concrete events in the clip rather than generic alternatives; and (3)~\emph{tone consistency}, where all options use the same register and hedging level. This process yields 2{,}400 candidate MCQs. We remove 5.8\% that contain invalid option formats, repeated options, or near-duplicate questions within the same clip, leaving 2{,}260 questions. Full prompts and templates are provided in Appendix~\ref{app:construction_details}.

\paragraph{Video-agnostic shortcut filtering.}
Following adversarial filtering approaches that reduce dataset biases \citep{LeBras2020AdversarialFO,Sakaguchi2019AnAW}, we test each question under a \textsc{Stem+Options} condition. GLM-5V and Claude-Haiku-4.5 receive only the question stem and four options, without video frames, ASR transcript, or structured evidence. Items answered correctly by both validators are treated as likely containing option-level shortcuts that allow answering without interaction evidence, and are removed. This filter reduces, but cannot eliminate, surface-level answer cues. It removes 10.6\% of the remaining questions, leaving 2{,}020 candidates.

\paragraph{Evidence-supported diagnostic filtering.}
The two validators then inspect each question against the audited structured evidence and answer it under the full input condition. A question is removed if its gold answer is not explicitly supported by the evidence, if more than one option is compatible with the evidence, or if the question requires information outside the clip. A correct answer from both validators does not itself trigger removal. Evidence-supported questions are regenerated only when the validators also identify evidence leakage or weak distractor discrimination. This step is used to calibrate diagnostic difficulty rather than to construct a validator-hard subset. In the final set, neither validator answers 293 items correctly, exactly one answers 587 correctly, and both answer 442 correctly. Together, this filtering stage removes 34.4\% of the remaining candidates, leaving 1{,}326 questions.

\paragraph{Expert audit.}
Three expert auditors with backgrounds in education and second-language pedagogy independently review a 360-question stratified audit set comprising 300 random questions balanced across all layers and subtasks, and 60 high-risk questions selected from inference-heavy subtasks and cases where the two validators disagreed.
Each item is judged on four dimensions: question clarity, gold-answer correctness, evidence sufficiency, and subtask alignment.
Of the audited questions, 340 are accepted without change, 16 retained after minor revision, and 4 removed, yielding a 94.4\% direct acceptance rate and a 1.1\% removal rate.
Sampling adequacy analysis and inter-annotator agreement are reported in Appendix~\ref{app:expert_verification}.
After expert adjudication, 1{,}322 MCQs are retained in the final benchmark.

\subsection{Dataset Overview}
\label{sec:statistics}

\begin{figure}[b]
\centering
\includegraphics[width=0.75\columnwidth]{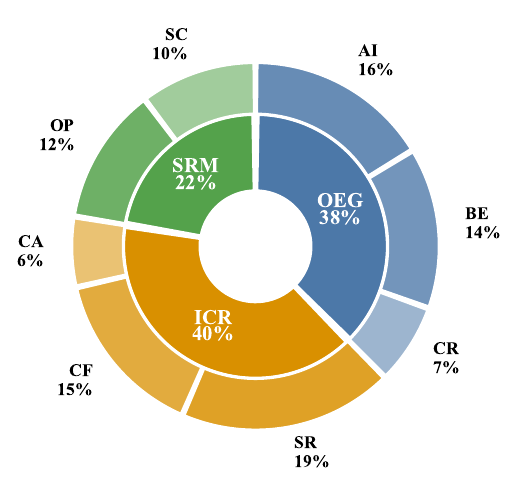}
\caption{Distribution of 1{,}322 MCQs across three diagnostic layers and eight subtasks.}
\label{fig:task_distribution}
\end{figure}

VIRA-Bench contains 1{,}322 multiple-choice questions over 1{,}005 clips, distributed across the three diagnostic layers as \oeg{} 497 questions, \icr{} 529 questions, and \srm{} 296 questions (Figure~\ref{fig:task_distribution}).
Clip durations range from 15 to 60 seconds (mean 32.4\,s), each centered on a single intervention and the student response it elicits.
Each question has exactly four options with one gold answer; option length is controlled within 20\% of the mean to reduce surface cues.
All clips are drawn from 72 learners and 22 teachers, providing natural variation in interaction style and pedagogical approach within the shared one-to-one tutoring format.
The benchmark scale reflects a coverage--audit tradeoff: every retained question is tied to a localized interaction window and audited evidence, rather than intended to cover all pedagogical settings.
Full per-subtask distributions and cross-lingual characteristics are in Appendix~\ref{app:distribution}.

  \begin{table*}[t]
\centering\footnotesize
\setlength{\tabcolsep}{6pt}
\renewcommand{\arraystretch}{1.05}
\resizebox{\textwidth}{!}{%
\begin{tabular}{@{}l*{3}{c}c*{3}{c}c*{2}{c}cc@{}}
\toprule
& \multicolumn{4}{c}{\textbf{\OEG{}}} & \multicolumn{4}{c}{\textbf{\ICR{}}} & \multicolumn{3}{c}{\textbf{\SRM{}}} & \\
\cmidrule(lr){2-5} \cmidrule(lr){6-9} \cmidrule(lr){10-12}
\textbf{Model} & AI & BE & CR & \cellcolor{gray!20}\textbf{Avg} & SR & CA & CF & \cellcolor{gray!20}\textbf{Avg} & SC & OP & \cellcolor{gray!20}\textbf{Avg} & \cellcolor{gray!20}\textbf{All [95\% CI]} \\
\midrule
Human\textsuperscript{$\dagger$} & 92.0 & 88.0 & 96.0 & \cellcolor{gray!20}92.0 & 92.0 & 84.0 & 88.0 & \cellcolor{gray!20}88.0 & 92.0 & 96.0 & \cellcolor{gray!20}94.0 & \cellcolor{gray!20}91.0 [86.2, 94.2] \\
\midrule
\multicolumn{13}{c}{\textit{Closed-source}} \\[2pt]
Gemini 3 Pro & 67.4 & 58.0 & 67.0 & \cellcolor{gray!20}63.8 & 64.0 & 51.2 & 52.3 & \cellcolor{gray!20}57.7 & 67.6 & 60.6 & \cellcolor{gray!20}63.9 & \cellcolor{gray!20}61.3 [58.7, 63.9] \\
Gemini 3 Flash & 58.6 & 61.7 & 70.2 & \cellcolor{gray!20}62.0 & 56.4 & 61.0 & 50.8 & \cellcolor{gray!20}55.0 & 57.4 & 70.0 & \cellcolor{gray!20}64.2 & \cellcolor{gray!20}59.7 [57.0, 62.3] \\
Qwen3.6-Plus & 54.9 & 51.6 & 62.8 & \cellcolor{gray!20}55.1 & 66.0 & 48.8 & 41.6 & \cellcolor{gray!20}54.3 & 51.5 & 67.5 & \cellcolor{gray!20}60.1 & \cellcolor{gray!20}55.9 [53.2, 58.6] \\
Qwen3.6-Flash & 48.8 & 36.7 & 46.8 & \cellcolor{gray!20}43.9 & 36.0 & 34.1 & 38.6 & \cellcolor{gray!20}36.7 & 40.4 & 46.2 & \cellcolor{gray!20}43.6 & \cellcolor{gray!20}40.9 [38.3, 43.6] \\
Qwen3.5-Flash & 24.7 & 18.6 & 22.3 & \cellcolor{gray!20}21.9 & 18.0 & 40.2 & 24.9 & \cellcolor{gray!20}24.0 & 41.9 & 20.6 & \cellcolor{gray!20}30.4 & \cellcolor{gray!20}24.7 [22.4, 27.1] \\
\midrule
\multicolumn{13}{c}{\textit{Open-source}} \\[2pt]
Kimi-K2.5 & 63.7 & 50.5 & 74.5 & \cellcolor{gray!20}60.8 & 53.2 & 53.7 & 44.7 & \cellcolor{gray!20}50.1 & 61.0 & 68.1 & \cellcolor{gray!20}64.9 & \cellcolor{gray!20}57.4 [54.7, 60.1] \\
MiMo-V2.5 & 54.0 & 62.8 & 56.4 & \cellcolor{gray!20}57.7 & 55.2 & 41.5 & 56.9 & \cellcolor{gray!20}53.7 & 72.8 & 51.2 & \cellcolor{gray!20}61.1 & \cellcolor{gray!20}56.9 [54.2, 59.5] \\
Qwen3-VL-235B-Think & 39.1 & 48.4 & 52.1 & \cellcolor{gray!20}45.1 & 39.2 & 42.7 & 38.1 & \cellcolor{gray!20}39.3 & 44.9 & 46.9 & \cellcolor{gray!20}45.9 & \cellcolor{gray!20}43.0 [40.3, 45.7] \\
Qwen3-VL-235B-Inst & 43.3 & 35.1 & 39.4 & \cellcolor{gray!20}39.4 & 42.8 & 29.3 & 26.4 & \cellcolor{gray!20}34.6 & 52.9 & 32.5 & \cellcolor{gray!20}41.9 & \cellcolor{gray!20}38.0 [35.5, 40.7] \\
Qwen3-VL-32B-Think & 17.7 & 26.6 & 19.1 & \cellcolor{gray!20}21.3 & 27.2 & 20.7 & 17.8 & \cellcolor{gray!20}22.7 & 20.6 & 23.8 & \cellcolor{gray!20}22.3 & \cellcolor{gray!20}22.1 [19.9, 24.4] \\
Qwen3-VL-32B-Inst & 22.3 & 18.1 & 14.9 & \cellcolor{gray!20}19.3 & 13.6 & 32.9 & 23.9 & \cellcolor{gray!20}20.4 & 31.6 & 13.8 & \cellcolor{gray!20}22.0 & \cellcolor{gray!20}20.3 [18.3, 22.6] \\
\bottomrule
\end{tabular}}%
\caption{Leaderboard results under the \full{} condition. Accuracy and 95\% Wilson intervals are reported. $\dagger$ The human reference uses a stratified 200-question subset, while every model is evaluated on all 1{,}322 questions.}
\label{tab:main}
\end{table*}

\section{Experiments}
\label{sec:experiments}

\subsection{Experimental Setup}

\paragraph{Models and data.}
We evaluate 11 multimodal large language models on 1{,}322 MCQs from 1{,}005
clips, each containing a complete 15--60 second IRF exchange.
The closed-source set includes Gemini 3 Pro (\texttt{gemini-3-pro-preview})~\citep{google2025gemini3pro}, Gemini 3 Flash (\texttt{gemini-3-flash-preview})~\citep{google2025gemini3flash}, Qwen3.6-Plus~\citep{alibabacloud2026qwen36plus}, Qwen3.6-Flash~\citep{alibabacloud2026qwen36flash}, and Qwen3.5-Flash~\citep{alibabacloud2026qwen35flash}.
The open-source set includes Kimi-K2.5~\citep{Bai2026KimiKV}, MiMo-V2.5~\citep{xiaomi2026mimov25}, and four Qwen3-VL variants---235B-A22B and 32B, each in Instruct and Thinking modes~\citep{qwen2025qwen3vl}.
All models use default inference settings with video input sampled at 1 frame per second.
A human reference based on three expert answerers independently answering a stratified 200-question subset (25 per subtask) with majority voting provides an upper reference point under the same input condition (details in Appendix~\ref{app:human_reference}; sampling adequacy in Appendix~\ref{app:expert_verification}).

\paragraph{Protocol and controls.}
Models are evaluated zero-shot under the default \full{} condition with video,
ASR, a question, and four labeled options. Option order is shuffled for every
model--item pair and the gold label is remapped. We parse the returned option
letter, using normalized-text matching only as fallback. For evidence-dependency
analysis, \textsc{ASR Text Only} removes video to isolate language-only
performance. \textsc{Visual Shuffle with Fixed ASR} retains all frames and the
original transcript but permutes VCE-detected visual segments, disrupting the
intervention--response--follow-up order and potentially introducing local
video--text mismatch. \textsc{Synchronized Video+ASR Shuffle} instead permutes
each segment with its aligned ASR span, preserving local cross-modal alignment.
These controls do not affect leaderboard ranking. We report overall,
layer-level, and subtask-level accuracy; full details are in
Appendix~\ref{app:inference_details}.

\subsection{Human--Model Gap and Layer Profiles}
\label{sec:results}

Table~\ref{tab:main} reports leaderboard results under the \full{} condition.
The best model, Gemini 3 Pro, reaches 61.3\% overall accuracy, leaving a
29.7-point gap below the 91.0\% human reference on the stratified subset.
The five leading models cluster between 55.9\% and 61.3\% with overlapping
confidence intervals, while the remaining models range from 20.3\% to 43.0\%.
We therefore focus on the broad human--model gap and recurring layer profiles
rather than fine-grained model ranks.

Across diagnostic layers, all top-tier models follow a consistent ordering of \SRM{}\,$>$\,\OEG{}\,$>$\,\ICR{}.
Among the top five models, the \OEG{}--\ICR{} gap ranges from 0.8 to 10.7 points.
This recurring profile, rather than any individual model comparison, motivates
the evidence-dependency analyses below.
The human reference follows the same layer ordering (\SRM{} 94.0\% $>$ \OEG{} 92.0\% $>$ \ICR{} 88.0\%) but at uniformly high accuracy. Its 95\% interval is 86.2--94.2\%, leaving its lower bound about 25 points above the best model. This subset result supports a robust human-model gap but should not be read as a precise human estimate for all 1{,}322 questions. The three below-chance models do not collapse onto a single answer position.
Their predicted labels span A--D, with each label accounting for 22.0--26.6\%
of outputs, and all 1{,}322 responses per model are parsed successfully.
Their scores therefore reflect incorrect item-level choices rather than invalid
responses or a dominant-label parsing artifact. Together, the leaderboard and
layer scores establish the first experimental claim: current models remain far
from the human reference, with the most consistent weakness appearing in
interaction contingency reasoning.

\subsection{Response-Format and Shortcut Controls}
\label{sec:format_controls}

The leaderboard could be inflated if models exploit labeled choices or static
question cues rather than interaction evidence. We test these alternatives with
two controls. Under free response, models answer without labeled choices and a
deterministic normalized-text matcher scores the output; it agrees with manual
judgments on 96.0\% of a 50-item audit. Under \textsc{Stem+Options}, models receive
the question and answer options but no video, ASR, description, or construction
metadata.

\begin{figure}[t]
\centering
\includegraphics[width=\columnwidth]{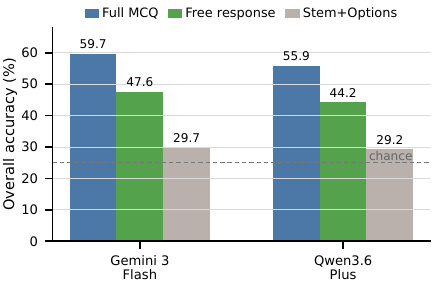}
\caption{Overall accuracy under full multiple-choice input, free response, and \textsc{Stem+Options}. The dashed line marks chance accuracy; layer-level scores are reported in Appendix~\ref{app:detailed_results}.}
\label{fig:response_controls}
\end{figure}

Figure~\ref{fig:response_controls} shows that free response lowers overall accuracy by 12.1 and 11.7 points, so
answer options provide useful scaffolding, yet the same layer ordering persists:
both models remain stronger on \OEG{} than \ICR{}. More decisively,
\textsc{Stem+Options} reaches only 29.2--29.7\%, compared with 55.9--59.7\%
under full Video+ASR input. Its layer scores are nearly flat, unlike the
layer-specific profiles under full input. Gold labels are balanced across A--D,
and all full-condition outputs are parsed successfully (Appendix~\ref{app:inference_details}).
Thus, response format affects absolute accuracy, but neither answer position nor
static question cues explains the main model separation or the \OEG{}--\ICR{}
pattern.

\subsection{Visual Evidence for OEG}
\label{sec:evidence_dependency}

We next test the source of each layer's evidence by evaluating five
representative models under the perturbation conditions defined in
\S\ref{sec:experiments}. Table~\ref{tab:ablation} first isolates what is lost when
the video is removed while the complete ASR transcript remains available.

\begin{table*}[t]
\centering\footnotesize
\setlength{\tabcolsep}{6pt}
\renewcommand{\arraystretch}{1.05}
\resizebox{\textwidth}{!}{%
\begin{tabular}{@{}ll*{3}{c}c*{3}{c}c*{2}{c}cc@{}}
\toprule
& & \multicolumn{4}{c}{\textbf{\OEG{}}} & \multicolumn{4}{c}{\textbf{\ICR{}}} & \multicolumn{3}{c}{\textbf{\SRM{}}} & \\
\cmidrule(lr){3-6} \cmidrule(lr){7-10} \cmidrule(lr){11-13}
\textbf{Model} & \textbf{Setting} & AI & BE & CR & \cellcolor{gray!20}\textbf{Avg} & SR & CA & CF & \cellcolor{gray!20}\textbf{Avg} & SC & OP & \cellcolor{gray!20}\textbf{Avg} & \cellcolor{gray!20}\textbf{All} \\
\midrule
\multirow{3}{*}{Gemini 3 Flash}
& Full & 58.6 & 61.7 & 70.2 & \cellcolor{gray!20}62.0 & 56.4 & 61.0 & 50.8 & \cellcolor{gray!20}55.0 & 57.4 & 70.0 & \cellcolor{gray!20}64.2 & \cellcolor{gray!20}59.7 \\
& ASR Text Only & 33.5 & 52.7 & 29.8 & \cellcolor{gray!20}40.0 & 48.4 & 46.3 & 48.2 & \cellcolor{gray!20}48.0 & 51.5 & 60.0 & \cellcolor{gray!20}56.1 & \cellcolor{gray!20}46.8 \\
& Visual Shuffle & 55.3 & 47.9 & 57.4 & \cellcolor{gray!20}52.9 & 32.8 & 34.1 & 40.6 & \cellcolor{gray!20}35.9 & 57.4 & 61.9 & \cellcolor{gray!20}59.8 & \cellcolor{gray!20}47.7 \\
\midrule
\multirow{3}{*}{Kimi-K2.5}
& Full & 63.7 & 50.5 & 74.5 & \cellcolor{gray!20}60.8 & 53.2 & 53.7 & 44.7 & \cellcolor{gray!20}50.1 & 61.0 & 68.1 & \cellcolor{gray!20}64.9 & \cellcolor{gray!20}57.4 \\
& ASR Text Only & 39.1 & 46.8 & 43.6 & \cellcolor{gray!20}42.9 & 52.8 & 50.0 & 38.6 & \cellcolor{gray!20}47.1 & 63.2 & 61.9 & \cellcolor{gray!20}62.5 & \cellcolor{gray!20}48.9 \\
& Visual Shuffle & 64.7 & 58.0 & 62.8 & \cellcolor{gray!20}61.8 & 37.6 & 35.4 & 36.5 & \cellcolor{gray!20}36.9 & 59.6 & 51.2 & \cellcolor{gray!20}55.1 & \cellcolor{gray!20}50.3 \\
\midrule
\multirow{3}{*}{MiMo-V2.5}
& Full & 54.0 & 62.8 & 56.4 & \cellcolor{gray!20}57.7 & 55.2 & 41.5 & 56.9 & \cellcolor{gray!20}53.7 & 72.8 & 51.2 & \cellcolor{gray!20}61.1 & \cellcolor{gray!20}56.9 \\
& ASR Text Only & 38.1 & 54.8 & 31.9 & \cellcolor{gray!20}43.3 & 44.0 & 41.5 & 47.2 & \cellcolor{gray!20}44.8 & 63.2 & 54.4 & \cellcolor{gray!20}58.4 & \cellcolor{gray!20}47.3 \\
& Visual Shuffle & 52.1 & 52.1 & 63.8 & \cellcolor{gray!20}54.3 & 30.0 & 40.2 & 41.6 & \cellcolor{gray!20}35.9 & 61.8 & 58.8 & \cellcolor{gray!20}60.1 & \cellcolor{gray!20}48.3 \\
\midrule
\multirow{3}{*}{Qwen3.6-Plus}
& Full & 54.9 & 51.6 & 62.8 & \cellcolor{gray!20}55.1 & 66.0 & 48.8 & 41.6 & \cellcolor{gray!20}54.3 & 51.5 & 67.5 & \cellcolor{gray!20}60.1 & \cellcolor{gray!20}55.9 \\
& ASR Text Only & 31.6 & 45.2 & 39.4 & \cellcolor{gray!20}38.2 & 54.0 & 47.6 & 44.7 & \cellcolor{gray!20}49.5 & 48.5 & 65.6 & \cellcolor{gray!20}57.8 & \cellcolor{gray!20}47.1 \\
& Visual Shuffle & 59.1 & 48.9 & 54.3 & \cellcolor{gray!20}54.3 & 35.6 & 34.1 & 39.6 & \cellcolor{gray!20}36.9 & 60.3 & 54.4 & \cellcolor{gray!20}57.1 & \cellcolor{gray!20}48.0 \\
\midrule
\multirow{3}{*}{Qwen3-VL-235B-Think}
& Full & 39.1 & 48.4 & 52.1 & \cellcolor{gray!20}45.1 & 39.2 & 42.7 & 38.1 & \cellcolor{gray!20}39.3 & 44.9 & 46.9 & \cellcolor{gray!20}45.9 & \cellcolor{gray!20}43.0 \\
& ASR Text Only & 32.6 & 34.6 & 27.7 & \cellcolor{gray!20}32.4 & 36.0 & 41.5 & 27.4 & \cellcolor{gray!20}33.6 & 45.6 & 44.4 & \cellcolor{gray!20}44.9 & \cellcolor{gray!20}35.7 \\
& Visual Shuffle & 42.3 & 43.1 & 44.7 & \cellcolor{gray!20}43.1 & 28.0 & 25.6 & 27.4 & \cellcolor{gray!20}27.4 & 47.8 & 42.5 & \cellcolor{gray!20}44.9 & \cellcolor{gray!20}37.2 \\
\bottomrule
\end{tabular}}%
\caption{Evidence dependency analysis on representative models. Subtask-level accuracy is reported under Full, ASR Text Only, and Visual Shuffle with Fixed ASR. Shaded columns show layer averages.}
\label{tab:ablation}
\end{table*}

Removing video reduces \OEG{} by 12.7--22.0 points across all five models.
Action Identification drops by 16--25 points and Content Recognition by 23--40
points across the four stronger models, whereas Behavior Extraction drops by
only 4--9 points. Teacher actions and screen content therefore require visual
grounding, while verbal cues partially support behavior extraction. ASR-only
overall accuracy remains 35.7--48.9\%, confirming that transcripts are
informative but cannot replace video. The intervention thus localizes the
incremental value of visual evidence to \OEG{}.
Paired McNemar tests sharpen this result: every \OEG{} decline under ASR-only
input remains significant after Holm correction, whereas the corresponding
\ICR{} and \SRM{} effects are smaller and less consistent. The no-video
intervention therefore removes information preferentially required for
observable grounding rather than imposing a uniform increase in difficulty.

\subsection{Temporal-Order Evidence for ICR}
\label{sec:temporal_dependency}

Table~\ref{tab:ablation} shows that \textsc{Visual Shuffle with Fixed ASR} causes the largest drops on \ICR{}, especially Sequence Recognition (16--30 points; e.g., Gemini 56.4$\to$32.8, MiMo 55.2$\to$30.0).
Causal Attribution and Contingent Feedback show model-dependent drops, indicating that the strongest temporal-order effect concentrates in Sequence Recognition.
Accuracy therefore declines sharply when the intervention--response--follow-up order is disrupted, even though the same visual frames remain available.
Across the five models, \ICR{} falls by 11.9--19.1 points, and all five declines
survive Holm correction. By contrast, four of five \OEG{} changes and four of
five \SRM{} changes are not significant under the same perturbation. This
selective pattern isolates order-bearing relations rather than uniform video
degradation
(Appendix~\ref{app:significance}).

\begin{figure}[t]
\centering
\includegraphics[width=\columnwidth]{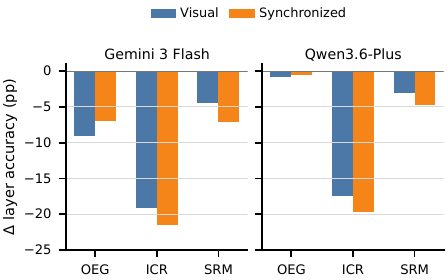}
\caption{Layer-level changes under fixed-ASR visual shuffle and synchronized video--ASR shuffle. The synchronized control reproduces the \ICR{}-selective degradation while preserving local cross-modal alignment.}
\label{fig:sync_control}
\end{figure}

Figure~\ref{fig:sync_control} shows that the synchronized control preserves local video--text alignment and yields the
same pattern. Relative to full input, \ICR{} falls by 21.5 points for Gemini 3
Flash and 19.7 points for Qwen3.6-Plus, while \OEG{} changes by 6.9 and 0.6
points. Cross-modal mismatch therefore cannot explain the \ICR{} decline by
itself.
The measured surface-complexity factors do not account for the \ICR{}-specific
effect: humans reach 88.0\%, experts accept 131/140 unchanged, stems are shortest
(13.0 words), option lengths match \OEG{}/\SRM{} (20.0 vs. 18.8/19.5 words), and
clip durations differ by at most 1.1 seconds across layers. Other unmeasured
differences among layers may remain.

\subsection{Diagnostic Implications}
\label{sec:implications}

The paired interventions turn layer scores into testable development targets.
For \OEG{}, the large no-video drop points to event-level alignment among speech,
teacher actions, and screen changes. For \ICR{}, the selective shuffle effect
identifies preservation of actor, target, and order across intervention,
response, and follow-up as the relevant evaluation target. For \SRM{}, the best
model remains 29.1 points below the human reference, motivating further tests of
learner-state tracking beyond partial cues from individual channels.

These targets also define how to interpret future gains. Stronger \OEG{} should
reflect better recovery of actions and screen state from video; stronger \ICR{}
should appear under intact order and diminish when the exchange is permuted; and
stronger \SRM{} should not be attributable to static question cues. Reporting
full-input scores together with perturbation deltas therefore distinguishes gains
in grounding, order-sensitive reasoning, and pedagogical judgment more clearly
than aggregate accuracy alone.

  \section{Conclusion}
\label{sec:conclusion}

We introduced VIRA-Bench, a focused benchmark that decomposes short-horizon pedagogical contingency understanding into three theory-anchored layers in one-to-one online tutoring. Across 11 models, the best reaches 61.3\% against a 91.0\% human reference on a stratified subset, and \ICR{} generally has the lowest layer accuracy among leading models. Removing video most strongly degrades \OEG{}, whereas disrupting event order most strongly degrades \ICR{}; this ICR-selective decline persists when aligned video--ASR segments are shuffled together. These results identify distinct behavioral evidence dependencies rather than an internal architectural mechanism. Within its one-to-one tutoring scope, VIRA-Bench separates progress in observable grounding from progress in order-sensitive contingency reasoning.

  \section{Limitations}
\label{sec:limitations}

VIRA-Bench focuses on 15--60-second excerpts from one-to-one online English tutoring, so its findings may not transfer directly to multi-party, cross-domain, or longer-horizon interactions.
Although its LLM-assisted construction is checked by multiple models and experts, occasional annotation artifacts may remain.
Finally, the multiple-choice video-plus-ASR protocol targets reproducible multimodal understanding rather than open-ended generation, action, or multi-turn adaptation.

  \section*{Ethical Considerations}

The source videos are online tutoring sessions involving child and adolescent learners, recorded by a commercial educational institution between 2018 and 2021 with parental or guardian consent for educational and research use.
The original data-collection protocol was approved by the source institution's ethics review committee.
The institution provided de-identified recordings under a data-sharing agreement limited to non-commercial academic research and prohibiting re-identification.
Before analysis or release, faces are blurred, identifying metadata are removed, and person mentions in transcripts are replaced with role placeholders; no question relies on identity, and the research team has no access to unblurred recordings or participant identities.
Only processed clips and transcripts are distributed under a restricted license, and VIRA-Bench is intended for model evaluation rather than high-stakes assessment of individual teachers or students.

  \section*{Acknowledgments}

This work is supported by the National Natural Science Foundation of China
(NSFC) under Grant No.~62672121 and the Major Project of the National Social
Science Foundation of China (NSSFC) under Grant No.~25\&ZD260. Peng Zhang and
Tun Lu are with the College of Computer Science and Artificial Intelligence,
Fudan University. Tun Lu is also affiliated with the Silver-X MOE Philosophy
\& Social Sciences Laboratory, the Fudan Institute on Aging, and the Shanghai
Key Laboratory of Data Science, Fudan University. We thank Jiahao Liu for
valuable feedback on this work.

  \bibliography{references}

  \appendix
  \section{Clip Extraction Details}
\label{app:extraction_params}

We extract clips from long online tutoring sessions through three stages.
First, ASR-guided coarse segmentation transcribes each session with Whisper-large-v3, identifies teacher utterances containing instructional functions (explanation, elicitation, correction, feedback, task prompting), and merges adjacent instructional intervals separated by less than 10 seconds.
This reduces each 1--2 hour session to approximately 15--30 minutes of likely interaction.

Second, Visual Content Event (VCE) detection samples shared-screen frames at 2\,fps within the ASR-selected intervals and compares consecutive frames using structural similarity \citep{Wang2004SSIM} and pixel-level difference ratio.
A VCE is triggered when either the SSIM score drops below 0.88 or the changed-pixel ratio exceeds 3\% of the frame area.
Events are classified by magnitude: \emph{large} changes (SSIM $<$ 0.88 and diff $>$ 30\%) indicate slide transitions or major screen rewrites, while \emph{local} changes (SSIM $<$ 0.88 and diff $>$ 3\%) capture teacher drawing, highlighting, or annotation.
A minimum gap of 2\,s is enforced between consecutive events to avoid duplicate triggers from gradual animations, and low-level cursor-only motion is removed unless accompanied by a stable pointing, drawing, highlighting, dragging, or slide-transition pattern.

Third, clip windows are generated by merging detected VCE events within a 5-second window into a single anchor group, then extending 3\,s before the first event and 12\,s after the last event.
The window is expanded to nearby ASR turn boundaries when this helps include the student response or teacher follow-up.
Clips are truncated at 60\,s and discarded if shorter than 15\,s.

\paragraph{Quality screening and face blurring.}
Candidate clips are screened by Gemini 2.5 Pro~\citep{google2025gemini25pro} on five dimensions: visual clarity, teaching activity, interaction completeness, audio quality, and benchmark value.
A clip is rejected if any dimension fails.
All faces are blurred using MediaPipe face detection with a $99{\times}99$ Gaussian kernel prior to any downstream processing; a 50-clip comparison before and after blurring confirmed accuracy differences below 3 percentage points, indicating that pedagogical signals reside primarily in the shared screen and speech.

\paragraph{Pipeline yield.}
Starting from 2{,}760 detected VCE anchors, anchor merging and duplicate removal yield 1{,}840 candidate clips.
Quality screening retains 1{,}480 clips, and structured evidence audit retains 1{,}370 clips.
After item-level filtering, 1{,}005 clips with at least one valid MCQ are kept in the final benchmark.

\begin{table}[t]
\centering\small
\resizebox{\columnwidth}{!}{%
\begin{tabular}{llr}
\toprule
\textbf{Stage} & \textbf{Parameter} & \textbf{Value} \\
\midrule
\multirow{4}{*}{VCE detection} & Sampling rate & 2\,fps \\
& SSIM threshold & 0.88 \\
& Changed-pixel ratio threshold & 3\% \\
& Minimum event gap & 2\,s \\
\midrule
\multirow{5}{*}{Clip generation} & Event merge window & 5\,s \\
& Pre-onset window & 3\,s \\
& Post-onset window & 12\,s \\
& Min clip duration & 15\,s \\
& Max clip duration & 60\,s \\
\midrule
\multirow{2}{*}{Face blurring} & Detection & MediaPipe \\
& Blur kernel & $99{\times}99$ Gaussian \\
\bottomrule
\end{tabular}}
\caption{Clip extraction pipeline parameters.}
\label{tab:extraction_params}
\end{table}

\section{Prompting and Item Construction Details}
\label{app:construction_details}

VIRA-Bench uses a construction-evaluation separation protocol.
GPT-4o drafts structured evidence descriptions, GPT-5 generates candidate MCQs and distractors, and GLM-5V and Claude-Haiku-4.5 serve as held-out validators.
None of these construction or validation models is included in the leaderboard.

\subsection{Structured Evidence Description}

GPT-4o receives the tutoring clip and its ASR transcript, and is instructed to output a structured JSON record.
The prompt specifies five evidence fields with typed sub-fields and timestamp requirements.
The model is instructed to focus on observable evidence and to mark uncertain observations explicitly rather than infer hidden mental states.

\begin{quote}\small
You are an expert educational video analyst specializing in online tutoring sessions. Your task is to produce a detailed, structured description of the teaching interaction in this video clip.

The video shows an online classroom with two main visual areas: (1) webcam feeds showing teacher and student (body language, gestures, and posture are visible); (2) teacher's shared screen showing instructional content.

Respond in JSON with the following fields:

\texttt{screen\_events}: list of \{timestamp\_start, timestamp\_end, action\_type, action\_description, content\_context, pedagogical\_tool\}. Action types: drawing\_writing, pointing\_highlighting, slide\_transition, drag\_drop\_interactive, animation\_reveal, none\_ambient.

\texttt{student\_behaviors}: list of \{timestamp\_start, timestamp\_end, behavior\_type, behavior\_description, verbal\_content, body\_language\}. Behavior types: verbal\_active, verbal\_passive, gestural\_physical, screen\_interaction, no\_observable\_response.

\texttt{teacher\_speech}: list of \{timestamp, content, speech\_act\}. Speech acts: instruction, question, feedback\_positive, feedback\_corrective, explanation, elicitation, social\_talk.

\texttt{temporal\_dynamics}: \{intervention\_response\_pattern, response\_latency\_seconds, causal\_chain\_description\}. Patterns: immediate ($<$3s), delayed (3--10s), cumulative, interleaved, no\_response.

\texttt{cognitive\_demand}: one of recall\_recognize, apply\_practice, analyze\_create.
\end{quote}

\subsection{MCQ Generation}

GPT-5 receives the audited evidence record, a target layer/subtask, and per-subtask instructions specifying the cognitive operation to test.
It generates one four-option MCQ per subtask in a single pass, including the stem, four options, the gold answer, and a distractor strategy label.

The system prompt enforces the following global constraints:

\begin{quote}\small
\textbf{Global Rules.}
(1) Each MCQ has exactly 4 options (A--D) with exactly one correct answer.
(2) Distractors must be plausible but clearly wrong given the video evidence.
(3) Anti-information-leakage: the question stem + options alone must not be sufficient to determine the correct answer without watching the video.
(4) Anti-shortcut: distribute correct answers roughly evenly across A--D; include distractors that would fool a model relying only on audio, only on vision, or only on common sense.
(5) Evidence-based: the correct answer must be verifiable from the structured description.
(6) Questions must require understanding the video, not just text.

\textbf{Option Length Parity.}
All four options must be approximately equal in length (within $\pm$20\% of the average option length in characters). The correct answer must not be systematically longer or more detailed than distractors.

\textbf{Distractor Quality.}
Distractors must share the same grammatical structure, level of specificity, and concreteness as the correct answer. Never use absolute language (``no'', ``never'', ``always'') or extreme scenarios. Never use hedging language that contrasts with concrete correct answers.

\textbf{Anti-Transcript-Leakage.}
The question stem must not quote or closely paraphrase specific speech content from the clip. The correct answer should depend on visual information that cannot be inferred from the audio transcript alone.
\end{quote}

Each subtask receives dedicated instructions specifying the target cognitive operation, the evidence fields to focus on, and the distractor design strategy. For example, the Action Identification (AI) subtask instructs the model to focus on \texttt{screen\_events} and generate distractors from plausible but non-occurring action types; the Causal Attribution (CA) subtask instructs the model to focus on \texttt{temporal\_dynamics} and generate distractors that reverse causal order or confuse co-occurrence with causation.

\subsection{Generate-Judge-Regenerate Loop}

Each clip undergoes up to three iterations of a closed-loop quality cycle:

\begin{enumerate}[nosep,leftmargin=*]
\item \textbf{Generate}: GPT-5 produces one MCQ per pending subtask.
\item \textbf{Judge}: Each MCQ passes through three sequential checks (described below). Failed MCQs receive structured feedback identifying the specific violation.
\item \textbf{Regenerate}: Failed MCQs are regenerated with the original prompt augmented by the judge feedback, which specifies what to fix.
\end{enumerate}

MCQs that pass all three checks are accepted; those that fail after three iterations are discarded.

\paragraph{Phase 1: Programmatic quality checks.}
Before any API call, each MCQ is checked for: (a)~option length ratio (max/min $\leq$ 1.2); (b)~absence of absolute language in distractors; (c)~valid JSON structure with exactly four options; (d)~correct answer present among options. Items failing these checks are immediately returned for regeneration with specific feedback.

\paragraph{Phase 2: Shortcut detection.}
GLM-5V and Claude-Haiku-4.5 receive only the question stem and four options, without video, ASR transcript, or evidence description. Each model selects an answer and reports confidence and any detected shortcut type (length bias, absolute language, semantic outlier, common sense, grammatical cue). If both validators answer correctly, the item is flagged as containing an option-level shortcut and returned for regeneration.

\begin{quote}\small
You are taking a multiple-choice test about educational video content. You have NOT seen any video, transcript, or context. You ONLY see the question stem and four options.

Select the answer you think is most likely correct. Rate your confidence: high, medium, or low. If you detect a shortcut (e.g., one option is obviously different, uses absolute language, or is much longer/shorter), describe it.
\end{quote}

\paragraph{Phase 3: Evidence-supported diagnostic filtering.}
GLM-5V and Claude-Haiku-4.5 receive the question stem, four options, and the full audited evidence description. If both validators answer correctly and judge the item as trivial or weakly discriminative for the intended subtask, the item is returned for regeneration. This step calibrates diagnostic difficulty without reference to any specific input modality.

\subsection{Item Filtering Statistics}

Table~\ref{tab:filtering_stats} summarizes the multi-stage filtering pipeline. After expert adjudication of the remaining items, the final benchmark contains 1{,}322 MCQs.

\begin{table}[t]
\centering\small
\resizebox{\columnwidth}{!}{%
\begin{tabular}{lrr}
\toprule
\textbf{Stage} & \textbf{Removed} & \textbf{Remaining} \\
\midrule
Initial generation (GPT-5) & --- & 2{,}400 \\
Format + duplicate checks & 5.8\% & 2{,}260 \\
Shortcut filtering (\textsc{Stem+Options}) & 10.6\% & 2{,}020 \\
Evidence-supported difficulty filtering & 34.4\% & 1{,}326 \\
Expert adjudication & 0.3\% & 1{,}322 \\
\bottomrule
\end{tabular}}
\caption{Item filtering pipeline yield. Percentages are relative to the input of each stage.}
\label{tab:filtering_stats}
\end{table}

\section{Expert Verification}
\label{app:expert_verification}

We conduct a single-round expert audit to estimate residual item errors after full-set automatic filtering.
Three domain experts independently review an audit set of 360 items (300 stratified random items covering all layers and subtasks, plus 60 high-risk items selected from inference-heavy subtasks and validator-disagreement cases).
Each item is judged along four dimensions: question clarity, gold-answer correctness, evidence sufficiency, and subtask alignment, each annotated as \textsc{Accept}, \textsc{Minor Revision}, or \textsc{Reject}.

\paragraph{Audit instructions.}
The auditors received the following instructions:
\begin{quote}\small
Review the complete de-identified video clip and its ASR transcript, followed by the question, four answer options, gold answer, and assigned subtask. Evaluate each item independently along four dimensions: question clarity, gold-answer correctness, evidence sufficiency, and subtask alignment. Mark \textsc{Accept} when the item is clear, has exactly one evidence-supported answer, and matches the assigned subtask. Mark \textsc{Minor Revision} when a limited wording, option-specificity, or evidence-target correction can resolve the issue without changing the intended reasoning. Mark \textsc{Reject} when the question is ambiguous, the gold answer is unsupported, multiple options are valid, the required evidence is absent from the clip, or the item does not match its assigned subtask. Base all judgments only on evidence available in the video and transcript.
\end{quote}

\paragraph{Recruitment and compensation.}
The three auditors were recruited based on their graduate-level training in applied linguistics or educational technology and their prior experience with classroom-interaction analysis. They were compensated at the institution's standard research-assistant rate for comparable expert annotation work.

Among the 360 audited items, 340 are accepted without change, 16 are retained after minor revision, and 4 are removed, yielding a 94.4\% direct acceptance rate and 98.9\% final acceptance rate.
For the stratified random split, 288/300 are accepted without change (2 removed); for the high-risk split, 52/60 are accepted without change (2 removed).
Among the 16 revised items, 7 involved tightening question wording, 5 involved rebalancing option specificity, and 4 involved clarifying the evidence target.
Among the 4 removed items, 2 had insufficient visual evidence, 1 had an ambiguous gold answer, and 1 had a subtask mismatch.

\paragraph{Agreement and stratified results.}
Fleiss' $\kappa$ on the binary acceptability decision is 0.76 (pairwise agreement 96.7\%).
Per-dimension $\kappa$: question clarity 0.82, gold-answer correctness 0.79, evidence sufficiency 0.68, subtask alignment 0.74.
Evidence sufficiency shows the lowest agreement, particularly for \srm{} items where experts differ on whether short clips provide sufficient context.
Table~\ref{tab:audit_stratified} reports audit outcomes by layer; acceptance rates are uniformly high, with \icr{} showing slightly more revisions due to the difficulty of judging temporal evidence sufficiency.

\begin{table}[t]
\centering\small
\resizebox{\columnwidth}{!}{%
\begin{tabular}{lccc}
\toprule
\textbf{Layer} & \textbf{Audited} & \textbf{Accept} & \textbf{Revise/Remove} \\
\midrule
\oegtab{} & 135 & 130 (96.3\%) & 4 / 1 \\
\icrtab{} & 140 & 131 (93.6\%) & 7 / 2 \\
\srmtab{} & 85 & 79 (92.9\%) & 5 / 1 \\
\midrule
Total & 360 & 340 (94.4\%) & 16 / 4 \\
\bottomrule
\end{tabular}}
\caption{Expert audit results stratified by layer. Accept = no change needed. Revise = minor wording adjustment. Remove = item excluded from final benchmark.}
\label{tab:audit_stratified}
\end{table}

\paragraph{Sampling adequacy.}
The 360-item audit set represents a 27\% sampling rate of the final benchmark.
At the observed 1.1\% removal rate (4/360), the 95\% Wilson confidence interval is [0.4\%, 2.8\%], bounding the expected number of residual defective items in the full 1{,}322-item set to fewer than 37.
This human audit is the final layer in a multi-stage quality pipeline: automated shortcut filtering removes 10.6\% of candidates, difficulty calibration removes trivially easy items, and model-based evidence validation removes 34.4\% of candidates before expert review.

\section{Benchmark Statistics}
\label{app:distribution}

\begin{table}[t]
\centering\small
\resizebox{\columnwidth}{!}{%
\begin{tabular}{llcc}
\toprule
\textbf{Layer} & \textbf{Subtask} & \textbf{MCQs} & \textbf{\%} \\
\midrule
\multirow{3}{*}{\oegtab{}} & AI (Action Identification) & 215 & 16.3 \\
& BE (Behavior Extraction) & 188 & 14.2 \\
& CR (Content Recognition) & 94 & 7.1 \\
\midrule
\multirow{3}{*}{\icrtab{}} & SR (Sequence Recognition) & 250 & 18.9 \\
& CA (Causal Attribution) & 82 & 6.2 \\
& CF (Contingent Feedback) & 197 & 14.9 \\
\midrule
\multirow{2}{*}{\srmtab{}} & SC (Support Calibration) & 136 & 10.3 \\
& OP (Outcome Prediction) & 160 & 12.1 \\
\midrule
\multicolumn{2}{l}{\textbf{Total}} & \textbf{1,322} & 100.0 \\
\bottomrule
\end{tabular}}
\caption{MCQ distribution across subtasks.}
\label{tab:distribution}
\end{table}

\paragraph{Clip and item statistics.}
Across 1{,}005 clips, the mean duration is 32.4\,s (SD = 11.4\,s; 25th/50th/75th percentiles: 23/30/40\,s; 94\% within the 15--60\,s design window).
Each clip has between 1 and 3 MCQs (mean 1.32): 72\% have exactly 1, 24\% have 2, and 4\% have 3. Multiple MCQs on the same clip always target different subtasks.

\paragraph{Cross-lingual characteristics.}
\label{app:crosslingual}
All source sessions are online English tutoring, where the teacher instructs in English while adapting explanations to students with diverse first-language backgrounds (primarily Mandarin Chinese, with some English L1 students). Teacher speech naturally includes code-switching between English instruction and native-language clarification. The shared screen content is predominantly in English (instructional materials, exercises, vocabulary).

The source corpus contains over 1{,}000 recorded sessions (each 1--2 hours); the 1{,}005 clips are drawn from sessions selected for instructional richness and interaction density. Students include both Chinese L1 and English L1 learners at primary and secondary levels. Teacher speech is approximately 75\% English and 25\% other languages by word count (estimated from ASR output). ASR is performed with Whisper-large-v3, which handles code-switched multilingual input.

Because all sessions share the same instructional format (English-medium online tutoring), the primary variation is in student L1 background and teacher code-switching patterns rather than in the language of instruction itself. We acknowledge the limited diversity of instructional contexts as a scope limitation in \S\ref{sec:limitations}.

\section{Evaluation Protocol}
\label{app:inference_details}

All models use provider-default generation parameters (temperature and top-$p$ are not overridden). Video clips are uploaded at native resolution (720p, 15--60\,s duration) and sampled at 1 frame per second across all providers. The ASR transcript is provided as a separate text field in the prompt. All models are queried via their respective provider APIs in a zero-shot setting with no system prompt.

\paragraph{Inference prompt templates.}
Under the default \full{} condition, models receive the video clip together with the ASR transcript provided as a separate text input. The transcript is not burned into the video frames; models read it from the prompt text. The prompt is:

\begin{quote}\small
Watch the video clip and read the speech transcript, then answer the question.\\
Use the transcript as spoken-language evidence; use the video for visual context.

\#\# Speech Transcript\\
\{transcript\}

\#\# Question\\
\{stem\}

\#\# Options\\
A. \{option\_A\}\\
B. \{option\_B\}\\
C. \{option\_C\}\\
D. \{option\_D\}

Return ONLY valid JSON: \{"answer": "<A/B/C/D>", "confidence": "high$|$medium$|$low", "reasoning": "brief explanation"\}
\end{quote}

Under the \textsc{ASR Text Only} condition, the video is removed entirely and only the speech transcript is provided:

\begin{quote}\small
Read the speech transcript from a tutoring clip, then answer the question. You do NOT have access to the video.

\#\# Speech Transcript\\
\{transcript\}

\#\# Question\\
\{stem\}

\#\# Options\\
A. \{option\_A\} \ldots D. \{option\_D\}

Return ONLY valid JSON: \{"answer": "<A/B/C/D>", "confidence": \ldots\}
\end{quote}

Under the \textsc{Visual Shuffle with Fixed ASR} condition, the video segments are permuted at VCE-detected scene boundaries before upload while the original transcript remains fixed. The prompt is identical to the \full{} condition; models are not informed that the temporal order has been altered.

Under the \textsc{Synchronized Video+ASR Shuffle} condition, each visual segment is paired with its timestamp-aligned ASR span and the pairs are permuted together. Order within each pair is preserved. This retains local video-text alignment while disrupting the original intervention, response, and follow-up sequence.

Under the \textsc{Static Frame} condition, 5 evenly-spaced key frames are extracted from the clip and presented as images:

\begin{quote}\small
Look at these 5 key frames extracted from a video clip of an online tutoring session, then answer the question.

\#\# Question\\
\{stem\}

\#\# Options\\
A. \{option\_A\} \ldots D. \{option\_D\}

Return ONLY valid JSON: \{"answer": "<A/B/C/D>", "confidence": \ldots\}
\end{quote}

\paragraph{Response parsing and scoring.}
Models are instructed to return a JSON object containing an answer letter (A--D), a confidence level, and a brief reasoning string. We first attempt JSON extraction via regex; if parsing fails, we fall back to extracting a bare option letter or matching the normalized response to the option text. Responses that match neither pattern are scored as incorrect. After fallback parsing, all models have zero invalid outputs under the full condition.
Option order is randomized independently for each model--item pair using a fixed random seed, and the gold label is remapped accordingly.
For API-based models, responses indicating the model did not receive the video trigger up to 3 automatic retries before being marked as failures.
Accuracy is computed as the proportion of items matching the gold answer (no partial credit; random baseline 25\%).
We report accuracy at three levels: overall (1{,}322 items), layer-level (\oeg{}, \icr{}, \srm{}), and subtask-level (eight subtasks), computed directly over items in each group.

\paragraph{Answer-label diagnostics.}
The remapped gold labels are balanced with 330 A labels, 331 B labels, 331 C labels, and 330 D labels. Table~\ref{tab:output_labels} reports model output labels under the full condition. The distributions show no output collapse and zero invalid responses. On the 924 errors shared by the two 32B Qwen3-VL variants, the models choose the same distractor content in 47.6\% of cases after independent option shuffling. A permutation test gives $p<0.001$, which is consistent with related models favoring the same plausible distractors rather than a fixed answer position.

\begin{table*}[t]
\centering\small
\begin{tabular}{lrrrrrr}
\toprule
\textbf{Model} & \textbf{Accuracy} & \textbf{A} & \textbf{B} & \textbf{C} & \textbf{D} & \textbf{Invalid} \\
\midrule
Gemini 3 Pro & 61.3 & 293 & 345 & 342 & 342 & 0 \\
Gemini 3 Flash & 59.7 & 337 & 320 & 317 & 348 & 0 \\
Kimi-K2.5 & 57.4 & 291 & 348 & 344 & 339 & 0 \\
MiMo-V2.5 & 56.9 & 291 & 353 & 336 & 342 & 0 \\
Qwen3.6-Plus & 55.9 & 328 & 332 & 330 & 332 & 0 \\
Qwen3-VL-235B-Think & 43.0 & 307 & 334 & 343 & 338 & 0 \\
Qwen3.6-Flash & 40.9 & 295 & 340 & 338 & 349 & 0 \\
Qwen3-VL-235B-Inst & 38.0 & 302 & 352 & 331 & 337 & 0 \\
Qwen3.5-Flash & 24.7 & 291 & 340 & 340 & 351 & 0 \\
Qwen3-VL-32B-Think & 22.1 & 326 & 324 & 338 & 334 & 0 \\
Qwen3-VL-32B-Inst & 20.3 & 332 & 334 & 319 & 337 & 0 \\
\bottomrule
\end{tabular}
\caption{Model output-label distributions under the full condition. Accuracy is reported in percent and label columns report counts.}
\label{tab:output_labels}
\end{table*}

\paragraph{Statistical testing.}
For the evidence dependency analysis, we apply McNemar's test for paired binary outcomes, comparing item-level accuracy under the \full{} condition against each perturbation condition at the layer level. We apply Holm correction for multiple comparisons and report adjusted $p$-values with significance thresholds at $\alpha = 0.05$, $0.01$, and $0.001$.

\section{Human Reference Details}
\label{app:human_reference}

The 200-question human evaluation set is stratified by subtask (25 questions $\times$ 8 subtasks), sampled uniformly at random from the full 1{,}322-question set.
Three domain experts with graduate training in applied linguistics independently answered all 200 questions after watching the corresponding video clips with ASR transcripts.
Table~\ref{tab:human_detail} reports individual and majority-vote accuracy.

\begin{table}[t]
\centering\footnotesize
\resizebox{\columnwidth}{!}{%
\begin{tabular}{lcccc}
\toprule
& \textbf{Exp.\,1} & \textbf{Exp.\,2} & \textbf{Exp.\,3} & \textbf{Majority} \\
\midrule
\oegtab{} (75 Qs) & 90.7 & 89.3 & 92.0 & 92.0 \\
\icrtab{} (75 Qs) & 84.0 & 80.0 & 84.0 & 88.0 \\
\srmtab{} (50 Qs) & 88.0 & 86.0 & 90.0 & 94.0 \\
\midrule
Overall (200 Qs) & 87.5 & 85.0 & 88.5 & 91.0 \\
\bottomrule
\end{tabular}}
\caption{Human reference accuracy (\%) by layer and expert. Majority = majority vote across three experts.}
\label{tab:human_detail}
\end{table}

Using the Wilson score interval at 95\% confidence:
\oeg{}: 92.0\% [83.6\%, 96.3\%];
\icr{}: 88.0\% [78.7\%, 93.6\%];
\srm{}: 94.0\% [83.8\%, 97.9\%];
Overall: 91.0\% [86.2\%, 94.2\%].
The lower bound of the overall CI (86.2\%) remains substantially above the best model (61.3\%), confirming the human--model gap is robust.
Individual expert totals are 175, 170, and 177 out of 200 (mean 87.0\%); majority voting raises accuracy to 91.0\%, indicating that disagreements are resolved by at least two experts agreeing.
Pairwise agreement on the binary correct/incorrect decision exceeds 90\% across all expert pairs, with agreement highest on \oeg{} items and lowest on \icr{} Causal Attribution items.
This sample size is comparable to human evaluation protocols in recent video understanding benchmarks \citep{Li2023MVBenchAC,Fu2024VideoMMETF}.

\section{Paired Significance Analysis of Evidence Dependency}
\label{app:significance}

To determine whether the observed performance changes are statistically reliable, we conduct two sided McNemar tests using paired item level correctness outcomes.
For each model and diagnostic layer, we compare predictions under the \textbf{Full} setting against predictions under \textbf{ASR Text Only} and \textbf{Visual Shuffle with Fixed ASR} on the same set of questions.
ASR Text Only removes visual evidence while retaining the standardized ASR input.
Visual Shuffle with Fixed ASR retains the ASR input and visual content, but disrupts the temporal organization of the visual stream.
We apply Holm correction across all comparisons reported in Table~\ref{tab:significance}.

\begin{table*}[t]
\centering
\small
\setlength{\tabcolsep}{4.5pt}
\begin{tabular}{@{}ll rr rr rr@{}}
\toprule
& & \multicolumn{2}{c}{\textbf{\oegtab{}}} & \multicolumn{2}{c}{\textbf{\icrtab{}}} & \multicolumn{2}{c}{\textbf{\srmtab{}}} \\
\cmidrule(lr){3-4} \cmidrule(lr){5-6} \cmidrule(lr){7-8}
\textbf{Model} & \textbf{Condition} & $\Delta$ & $p_{\mathrm{adj}}$ & $\Delta$ & $p_{\mathrm{adj}}$ & $\Delta$ & $p_{\mathrm{adj}}$ \\
\midrule
\multirow{2}{*}{Gemini 3 Flash}
& ASR Text Only    & $-$21.9 & ${<}.001$*** & $-$7.0  & ${<}.001$*** & $-$8.1 & ${<}.01$** \\
& Visual Shuffle & $-$9.1  & ${<}.001$*** & $-$19.1 & ${<}.001$*** & $-$4.4 & n.s. \\
\midrule
\multirow{2}{*}{Kimi-K2.5}
& ASR Text Only    & $-$17.9 & ${<}.001$*** & $-$3.0  & n.s.         & $-$2.4 & n.s. \\
& Visual Shuffle & $+$1.0  & n.s.         & $-$13.2 & ${<}.001$*** & $-$9.8 & ${<}.001$*** \\
\midrule
\multirow{2}{*}{MiMo-V2.5}
& ASR Text Only    & $-$14.5 & ${<}.001$*** & $-$8.9  & ${<}.001$*** & $-$2.7 & n.s. \\
& Visual Shuffle & $-$3.4  & n.s.         & $-$17.8 & ${<}.001$*** & $-$1.0 & n.s. \\
\midrule
\multirow{2}{*}{Qwen3.6-Plus}
& ASR Text Only    & $-$16.9 & ${<}.001$*** & $-$4.7  & ${<}.05$*    & $-$2.4 & n.s. \\
& Visual Shuffle & $-$0.8  & n.s.         & $-$17.4 & ${<}.001$*** & $-$3.0 & n.s. \\
\midrule
\multirow{2}{*}{Qwen3-VL-235B-Thinking}
& ASR Text Only    & $-$12.7 & ${<}.001$*** & $-$5.7  & ${<}.01$**   & $-$1.0 & n.s. \\
& Visual Shuffle & $-$2.0  & n.s.         & $-$11.9 & ${<}.001$*** & $-$1.0 & n.s. \\
\bottomrule
\end{tabular}
\caption{Paired significance analysis of evidence dependency. $\Delta$ denotes the change in accuracy, measured in percentage points, relative to the Full setting. Negative values indicate performance degradation. Adjusted $p$ values are obtained from two sided McNemar tests over paired item level correctness outcomes with Holm correction for multiple comparisons. \textit{n.s.} indicates that the change is not significant after correction. *$p_{\mathrm{adj}}{<}.05$, **$p_{\mathrm{adj}}{<}.01$, ***$p_{\mathrm{adj}}{<}.001$.}
\label{tab:significance}
\end{table*}

The significance analysis supports a selective evidence dependency pattern across diagnostic layers.
Removing visual input produces the largest degradation on \OEG{} for every evaluated model, with drops ranging from 12.7 to 21.9 percentage points.
All \OEG{} declines under ASR Text Only remain significant after correction.
This result indicates that observable teacher actions, learner behaviors, and shared screen content cannot be reliably recovered from ASR input alone.

Visual Shuffle with Fixed ASR produces a complementary effect.
Its largest degradation consistently occurs on \ICR{}, with drops ranging from 11.9 to 19.1 percentage points, all significant after correction.
Because the ASR input remains unchanged in this condition, the \ICR{} decline reflects the loss of visual temporal organization required to connect teacher intervention, learner response, and teacher follow up.
In contrast, Visual Shuffle with Fixed ASR has no significant adverse effect on \OEG{} for four of the five evaluated models, suggesting that the perturbation largely preserves access to local observable evidence while disrupting relation bearing temporal structure.

\SRM{} exhibits a less uniform response to the two perturbations.
Most \SRM{} changes are not significant after correction, with the exception of larger model specific effects such as the degradation under ASR Text Only for Gemini 3 Flash and under Visual Shuffle with Fixed ASR for Kimi-K2.5.
This mixed pattern is consistent with the design of \SRM{}, which requires predictive or evaluative judgments supported by multiple forms of evidence, including visual context, language cues, learner state, and broader instructional priors.

\paragraph{Effect of temporal shuffling.}
Visual Shuffle with Fixed ASR is constructed by permuting temporally coherent event segments while preserving the internal order of frames within each segment.
It therefore disrupts the ordering among intervention, response, and follow up without reducing each segment to visually incoherent frames.
The resulting performance pattern further indicates that the primary effect is not generic visual corruption.
\OEG{} remains largely stable for most models, whereas \ICR{} declines substantially and consistently.
At the subtask level, the strongest degradation is expected on Sequence Recognition, followed by relation focused tasks such as Causal Attribution Plausibility and Contingent Feedback Recognition.
Together, these results show that VIRA-Bench requires models to use visual temporal evidence to recover contingent pedagogical relations rather than merely recognize isolated events or rely on language cues.

\section{Detailed Control Results}
\label{app:detailed_results}

\begin{table}[t]
\centering\small
\setlength{\tabcolsep}{4pt}
\resizebox{\columnwidth}{!}{%
\begin{tabular}{llrrrr}
\toprule
\textbf{Model} & \textbf{Setting} & \textbf{OEG} & \textbf{ICR} & \textbf{SRM} & \textbf{All} \\
\midrule
\multirow{3}{*}{Gemini 3 Flash}
& Full MCQ & 62.0 & 55.0 & 64.2 & 59.7 \\
& Free response & 50.5 & 41.8 & 53.0 & 47.6 \\
& \textsc{Stem+Options} & 29.8 & 29.5 & 29.7 & 29.7 \\
\midrule
\multirow{3}{*}{Qwen3.6-Plus}
& Full MCQ & 55.1 & 54.3 & 60.1 & 55.9 \\
& Free response & 43.9 & 41.6 & 49.3 & 44.2 \\
& \textsc{Stem+Options} & 29.0 & 29.3 & 29.4 & 29.2 \\
\bottomrule
\end{tabular}}
\caption{Response-format and no-evidence controls. Accuracy is reported in percent.}
\label{tab:format_controls_detail}
\end{table}

\section{Positioning Against Causal Video QA}
\label{app:causal_positioning}

\begin{table}[t]
\centering\scriptsize
\setlength{\tabcolsep}{2.5pt}
\resizebox{\columnwidth}{!}{%
\begin{tabular}{l cccc}
\toprule
& \textbf{Causal} & \textbf{Participant} & \textbf{Theory-} & \textbf{Evidence} \\
& \textbf{reasoning} & \textbf{interaction} & \textbf{grounded} & \textbf{control} \\
\midrule
NExT-QA \citep{Xiao2021NExTQANP} & \cm & \xm & \xm & \xm \\
STAR \citep{Wu2024STARAB} & \cm & \xm & \xm & \xm \\
CLEVRER \citep{Yi2019CLEVRERCE} & \cm & \xm & \xm & \cm \\
Causal-VidQA \citep{Li2022FromRT} & \cm & \xm & \xm & \xm \\
Social-IQ 2.0 \citep{Zadeh2019SocialIQAQ} & \cm & \cm & \xm & \xm \\
\midrule
\textbf{VIRA-Bench} & \cm & \cm & \cm & \cm \\
\bottomrule
\end{tabular}}
\caption{Comparison with causal video QA benchmarks. VIRA-Bench uniquely combines causal reasoning over participant interaction with theory-grounded diagnostic layers and controlled evidence perturbation.}
\label{tab:causal_comparison}
\end{table}

VIRA-Bench differs from existing causal video QA benchmarks in three respects.
First, causal video QA benchmarks \citep{Xiao2021NExTQANP,Wu2024STARAB,Yi2019CLEVRERCE,Li2022FromRT} primarily test physical causality or event-level ``why'' questions in third-person observation settings. VIRA-Bench tests \emph{interactional} causality: whether one participant's action triggers another's response and shapes subsequent behavior.
Second, VIRA-Bench grounds its diagnostic layers in established pedagogical theory (IRF structure, contingent teaching), providing a principled decomposition rather than ad-hoc question categories.
Third, VIRA-Bench includes controlled evidence perturbation (ASR-only, temporal shuffle) as part of the evaluation protocol, enabling construct validity analysis beyond aggregate accuracy.

\section{Transcript De-identification}
\label{app:transcript_deidentification}

Spoken names can remain in ASR transcripts after visual and metadata de-identification. We therefore process all 1{,}005 released transcripts with multilingual named-entity recognition and rules for vocatives and honorific-plus-name constructions in English and Mandarin. The process detects 1{,}340 person mentions. Each mention is replaced with \texttt{[TEACHER]}, \texttt{[STUDENT]}, or \texttt{[SPEAKER]} when the role cannot be determined. Replacement rather than deletion preserves who addresses whom, which can be relevant to Sequence Recognition and Causal Attribution. The released transcripts are de-identified versions of the raw Whisper-large-v3 output used during construction. No subtask asks for identity and no name is used as answer-bearing evidence.

\section{Datasheet}
\label{app:datasheet}

We follow the Datasheet for Datasets framework \citep{Gebru2018DatasheetsFD}.

\paragraph{Motivation.}
VIRA-Bench was created to evaluate whether video LLMs can understand pedagogical interaction contingency in educational videos. Existing video QA benchmarks test physical causality or social perception but do not assess whether models can track the causal chain from teacher intervention through student response to instructional follow-up. The benchmark was created by the authors at their respective institutions.

\paragraph{Composition.}
The benchmark contains 1{,}322 four-option multiple-choice questions spanning 1{,}005 video clips from online tutoring sessions (duration statistics in Appendix~\ref{app:distribution}). Questions are distributed across three diagnostic layers (\oeg{}: 497, \icr{}: 529, \srm{}: 296) and eight subtasks. All clips depict real instructional interactions; no synthetic or staged content is included. The dataset does not contain offensive content. All faces are blurred.

\paragraph{Collection process.}
Source videos are online tutoring sessions recorded by a commercial educational institution between 2018 and 2021. Sessions involve English-medium instruction with Chinese L1 students at primary and secondary levels. Informed consent for educational and research use was obtained from all participants' parents or legal guardians. The authors selected sessions from a corpus of over 1{,}000 recorded sessions based on instructional richness and interaction density.

\paragraph{Preprocessing.}
All faces are blurred prior to any downstream processing. Clips are extracted via the VCE-based pipeline described in Appendix~\ref{app:extraction_params}. ASR transcription uses Whisper-large-v3. Released transcripts are de-identified as described in Appendix~\ref{app:transcript_deidentification}.

\paragraph{Uses.}
VIRA-Bench is intended for evaluating multimodal language models on pedagogical interaction understanding. It should not be used for teacher evaluation, student assessment, or any high-stakes educational decision-making. The benchmark is not designed for training; using it as training data would compromise its diagnostic validity.

\paragraph{Distribution.}
The benchmark will be released under a research-only license. Video clips are distributed as de-identified MP4 files. MCQ metadata (stems, options, gold answers, layer/subtask labels) are distributed as JSON. Raw tutoring sessions are not released. Access requires agreeing to terms prohibiting re-identification attempts and commercial use.

\paragraph{Maintenance.}
The benchmark is maintained by the authors. Version updates will be documented in the release repository. We do not plan scheduled updates but will issue corrections if errors are identified post-release.

\end{document}